\documentclass{Rinton-P9x6}

\DeclareMathAlphabet   {\mathsc}{OT1}{cmr}{m}{sc}

\def\[{\left [}
\def\]{\right ]}
\def\({\left (}
\def\){\right )}
\newcommand{\lang}{\left\langle}
\newcommand{\rang}{\right\rangle}

\newcommand{\oline}[1]{\overline{#1}}

\newcommand{\wtd}[1]{\widetilde{#1}}

\newcommand{\TeV}      {~\mathrm{TeV}}

\newcommand{\STR}      {\mathsc{str}}

\newcommand{\lowest}{|_{\theta =\bar{\theta}=0}}

\newcommand{\order}{\mathcal{O}}
\newcommand{\re}{{\rm Re}}

\newcommand{\gappeq}{\mathrel{\rlap {\raise.5ex\hbox{$>$}}
{\lower.5ex\hbox{$\sim$}}}}
\newcommand{\lappeq}{\mathrel{\rlap{\raise.5ex\hbox{$<$}}
{\lower.5ex\hbox{$\sim$}}}}

\begin{document}

\title{Experimental Signatures of K\"ahler Stabilization of the Dilaton}

\author{Brent D. Nelson}

\address{Michigan Center for Theoretical Physics, \\ University
of Michigan, Ann Arbor, MI 48109, USA
}




\maketitle

\abstracts{We investigate the collider signatures of modular
invariant gaugino condensation, with K\"ahler stabilization of the
dilaton, in the context of weakly coupled heterotic string-based
models as an example of how supergravity can be used to build a
meaningful string phenomenology.}

\section{Background of the Model}
\label{sec:model}

The dilaton is the only one of the various possible string moduli
fields that always appears in the low-energy effective theory in a
uniform way. It represents the tree-level value of the gauge
kinetic function $f_a$ and thus its vacuum expectation value
determines the string coupling constant. In the chiral formulation
of the dilaton we have $f_a^{(0)} = S$ and  $<\re \; s>\; =
1/g_{\STR}^{2}$ where $s=S\lowest$ and $g_{\STR}$ is the universal
gauge coupling at the string scale.

It is clear that the low-energy phenomenology depends crucially on
finding a dynamical mechanism that ensures a finite vacuum value
for the dilaton at the observed coupling strength. However, the
superpotential for the dilaton is vanishing at the classical level
so only nonperturbative effects, of string and/or field-theoretic
origin, can create a superpotential capable of stabilizing the
dilaton.\cite{BaDi94} There are two commonly employed classes of
solutions to this challenge.\cite{DiSh01} The first, sometimes
referred to as the ``racetrack'' method, assumes only the tree
level form of the dilaton K\"ahler potential $K_{\rm
tree}(S,\oline{S}) = -\ln (S+\oline{S})$ and relies on at least
two gaugino condensates in the hidden sector to generate the
necessary dilaton superpotential. Generally the vacuum energy
remains nonzero in such scenarios. This method requires correctly
choosing the relative sizes of the beta-function coefficients for
two different condensing gauge groups.

The second approach, sometimes referred to as ``K\"ahler
stabilization,'' assumes that the tree level K\"ahler potential
for the dilaton is augmented by nonperturbative corrections of a
stringy or field-theoretic origin. Then in the presence of one or
more gaugino condensates in the hidden sector the dilaton can be
stabilized at $g_{\STR}^2 = 1/2$ with a vanishing vacuum energy.
This method requires correctly choosing parameters in the
postulated nonperturbative K\"ahler potential. This latter
approach gives rise to a scenario that Casas~\cite{Ca96} referred
to as the ``generalized dilaton-dominated'' scenario. The subject
of this talk is to consider this well-defined and well-motivated
model as a template for how supergravity effective theories can be
used to bridge the gap between string theory and experiment.

Consider the F-term scalar potential that arises from any generic
supergravity theory $ V= K_{I\bar{J}} F^I  \oline{F}^{\bar{J}} - M
\oline{M}/3$, where $F^I$ is the auxiliary field associated with
the chiral superfield $Z^I$ and $M$ is the auxiliary field of
supergravity. When only the dilaton auxiliary field $F^S$ receives
a vacuum expectation value the potential can be written
\begin{equation}
V = K_{s\bar{s}} |F^S|^{2} -3e^{K}|W|^{2} = e^{K} K^{s\bar{s}}
|W_s + K_sW|^{2} -3e^{K}|W|^2 . \label{dilpot}
\end{equation}
Requiring that the potential~(\ref{dilpot}) be vanishing in the
vacuum $\lang V\rang=0$ then implies (up to an overall phase)
\begin{equation}
F^{S} = \sqrt{3} m_{3/2} (K_{s\bar{s}})^{-1/2} =
\sqrt{3}m_{3/2}a_{\rm np}(K_{s\bar{s}}^{\rm tree})^{-1/2} ,
\label{FS}
\end{equation}
where we have introduced the parameter $a_{\rm np} \equiv
\(K_{s\bar{s}}^{\rm tree}/K_{s\bar{s}}^{\rm true}\)^{1/2}$
designed to measure the departure of the dilaton K\"ahler
potential from its tree level value due to nonperturbative
effects. Recall that $\lang (K_{s\bar{s}}^{\rm tree})^{1/2} \rang
= \lang 1/(s+\bar{s}) \rang = g_{\STR}^{2}/2 \simeq 1/4$.

To understand the likely magnitude of the phenomenological
parameter $a_{\rm np}$ let us make the quite well-grounded
assumption that the superpotential for the dilaton is generated by
the phenomenon of gaugino condensation and that its dilaton
dependence is given by $W(S) \propto e^{-3S/2b_{a}}$. Here $b_{a}$
is the beta-function coefficient of a condensing gauge group
${\cal G}_{a}$ of the hidden sector with $b_a = (1/16\pi^2)\(3 C_a
- \sum_i C_a^i\)$ and $C_a$, $C_a^i$ are the quadratic Casimir
operators for the gauge group ${\cal G}_a$, respectively, in the
adjoint representation and in the representation of the matter
fields $Z^i$ charged under that group. Let us assume a single
condensing gauge group, which we will denote by ${\cal G}_{+}$, so
that we can write $W_s = -(3/2b_{+})W(S)$.

Returning for a moment to the tree level case, it is not difficult
to see that requiring $\lang V \rang=0$ in~(\ref{dilpot}) would
require a dilaton {\em vev} such that $g_{\STR}^{2} \sim 1/b_{+}
\sim 16\pi^{2}$. In fact, no such minimum exists and the dilaton
has a runaway solution to zero coupling. However, if we do not
insist on the tree level dilaton K\"ahler potential then the
vanishing of the vacuum energy implies
\begin{equation}
(K_{s\bar{s}})^{-1}\left|K_s - \frac{3}{2b_{+}} \right|^{2} =3 \;
\; \to (K_{s\bar{s}})^{-1/2} = \sqrt{3}
\frac{\frac{2}{3}b_{+}}{1-\frac{2}{3}b_{+}K_{s}} . \label{Ktrue}
\end{equation}
So provided $K_s \sim \order(1)$ so that $K_s b_{+} \ll 1$ we can
immediately see that a K\"ahler potential which stabilizes the
dilaton while simultaneously providing zero vacuum energy will
{\em necessarily} result in a suppressed dilaton contribution to
soft supersymmetry breaking. Indeed, from the definition of
$a_{\rm np}$ we have
\begin{equation}
a_{\rm np}=\sqrt{3}\frac{\frac{2}{3}\frac{g_{s}^{2}}{2}b_{+}}{1-
\frac{2}{3}K_{s}b_{+}} \ll 1 . \label{aBGW}
\end{equation}

\section{Soft terms and benchmark choices} \label{sec:soft}

From~(\ref{FS}) we see that $|F^{S}/M| \simeq 4a_{\rm np}/\sqrt{3}
\ll 1$, so one-loop corrections can be important for those soft
supersymmetry-breaking terms that receive their tree level
contributions solely from the dilaton auxiliary field. In
particular, loop-corrections arising from the conformal anomaly
are proportional to $M$ itself and receive no suppression, so they
can be competitive with the tree level contributions in the
presence of a nontrivial K\"ahler potential for the dilaton. If we
assume that the K\"ahler metric for the observable sector matter
fields is independent of the dilaton then the leading order
expressions for the soft supersymmetry-breaking terms for
canonically normalized fields are
\begin{eqnarray}
M_{a} &\simeq& \frac{g_{a}^{2}(\mu)}{2}\[\lang F^S \rang
-2b_{a}m_{3/2}\] \nonumber \\ A_{ijk} &\simeq& -\lang K_s F^{S}
\rang + m_{3/2}\[\gamma_{i} + \gamma_{j} +\gamma_{k} \] \nonumber
\\ m_{0}^{2} &\simeq& m_{3/2}^{2} ,\label{softterms1}
\end{eqnarray}
where $\gamma_{i}$ is the anomalous dimension of field
$Z^{i}$.\cite{talk2,bench} While we have presented only the
leading terms in the one-loop parameters in~(\ref{softterms1}),
the complete expressions for soft terms\cite{BiGaNe01} at one loop
were used in the calculations.\footnote{The appropriate input
values for the {\tt PYTHIA} event generator can be obtained at
{\tt http://www-pat.fnal.gov/personal/mrenna/benchmarks/}.}

From~(\ref{softterms1}) it is clear that the dominant signature of
a ``generalized'' dilaton-domination scenario is the hierarchy
between gaugino and scalar masses. On top of this gross feature it
is also clear that the loop effects will produce a
``fine-structure'' of nonuniversalities among the gaugino masses
and A-terms. With the phase choice represented
in~(\ref{softterms1}), and the definition of $b_a$, the effect of
the loop corrections will be to lower the gluino mass $M_3$ while
increasing the bino mass $M_1$ relative to the wino $M_2$. It is
important to note that the significant splitting experienced by
the gaugino masses is not also seen in the gauge couplings
themselves. Tree level gaugino masses are still universal, but are
suppressed, so that nonuniversal loop contributions are
comparable, while loop contributions to the gauge couplings
themselves are always small in comparison to the large tree level
value.

In previous studies of this class of models\cite{talk1} it was
found that requiring $m_{3/2} \approx 1 \TeV$ to within an order
of magnitude typically required $b_{+} \leq 0.15$. We are thus led
to consider the cases where $b_+ = 15/16\pi^2 \simeq 0.095$ and
$b_+ = 9/16\pi^2 \simeq 0.057$. The former could result from a
condensation of pure $SU(5)$ Yang-Mills fields in the hidden
sector; the latter from a similar condensation of pure $SU(3)$
Yang-Mills fields or from the condensation of an $E_6$ hidden
sector gauge group with 9 $\mathbf{27}$'s condensing in the hidden
sector as well. To serve as a baseline we will also consider a
much larger value of $b_{+}= 36/16\pi^2 \simeq 0.228$. This could
result from a hidden sector condensation of pure $E_6$ Yang-Mills
fields.


\section{Phenomenological Features}

The resulting low-energy spectrum for the three benchmark models
is summarized in Table~\ref{tbl:spectraSUM}. Here the model
labeled ``mSUGRA'' is the cMSSM Point~B of Battaglia et
al.\cite{Ellis} The key feature of K\"ahler stabilization is
likely to be a small mass splitting between $\wtd{N}_1$ and
$\wtd{C}_1$ characteristic of anomaly mediation in the gaugino
sector. The LSP is not overwhelming bino-like nor is it a pure
wino state as in anomaly mediation. Thus adequate neutralino relic
density can be obtained.\cite{BiNe01} The relatively small $\mu$
term values for such large scalar masses is a manifestation of the
focus point effect.

\begin{table}[th]
\caption{Selected physical masses and parameters for benchmark
models.\label{tbl:spectraSUM}}
{\begin{center} \begin{footnotesize} \begin{tabular}{|l|c|c|c|c|}
\hline Point & A & B & C & mSUGRA
\\ \hline  $\tan\beta$ & 10 & 5 & 5 & 10
\\ $m_{3/2}$ & 1500 & 3200 & 4300 & NA
\\ $a_{\rm np}$ & 1/15.77 & 1/37.05 & 1/61.36 & NA
\\ \hline
$m_{\wtd{N}_1}$ & 77.9 & 93.1 & 90.6 & 98\\
$m_{\wtd{N}_{2}}$ & 122.3 & 132.2 & 110.0 & 182  \\
$m_{\wtd{C}_{1}^{\pm}}$ & 119.8 & 131.9 & 109.8 & 181  \\
$m_{\tilde{g}}$ & 471 & 427 & 329 & 582 \\
$\wtd{B} \; \% |_{\rm LSP}$ & 89.8 \% & 98.7 \% &  93.4 \% &
99.9\%\\
$\wtd{W}_{3} \% |_{\rm LSP}$ & 2.5 \% &  0.6 \% & 4.6 \% & 0.01\%
\\
\hline
$m_{\tilde{t}_{1}}$ & 947 & 1909 & 2570 & 392 \\
%
%
$m_{\tilde{b}_{1}}$ & 1282 & 2681 & 3614 & 501 \\
%
%
$m_{\tilde{\tau}_{1}}$ & 1491 & 3199 & 4298 & 137 \\ \hline
%
%
$m_{h}$ & 114.3 & 114.5 & 116.4  & 112 \\
$m_{A}$ & 1507 & 3318 & 4400 & 381 \\
$\mu$ & 245 & 631 & 481 & 332 \\ \hline
\end{tabular}
\end{footnotesize} \end{center}}
\end{table}

\begin{figure}[t]
\begin{center}
\epsfxsize=20pc 
\epsfbox{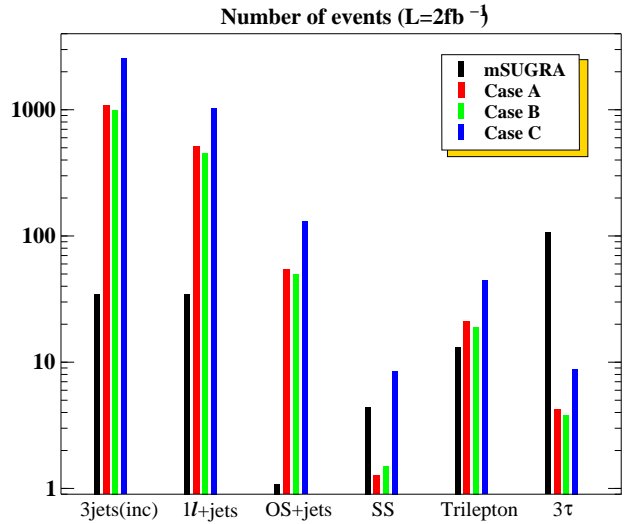} 
\caption{Number of superpartner events of different signatures at
the Tevatron with $2\mbox{fb}^{-1}$. \label{fig:bgwsig}}
\end{center}
\end{figure}

Figure~\ref{fig:bgwsig} shows naive estimates of numbers of events
in 2 fb$^{-1}$ integrated luminosity for various models and
various inclusive signatures.
The signature of these models are calculated using {\tt PYTHIA},
but only at the generator level: no geometric or kinematic cuts or
triggering efficiencies are applied, no jet clustering is
performed, tau leptons are not decayed, etc. The event numbers are
only meant to illustrate the generic features of each model and
demonstrate the experimental challenges. Every signature has
missing energy. From left to right, the signatures are: (1)
inclusive multi-jets $n_{jets}\ge 3$, (2) one lepton plus
$n_{jets}\ge 2$, (3) opposite sign dileptons plus $n_{jets}\ge 2$,
(4) same-sign dileptons, (5) trilepton, and (6) 3 taus plus jets
[before decaying the taus].
For signatures (4)-(6), no requirement is made on the number of
jets.
A background analysis must of course be done to be sure any given
channel is detectable, but models with hundreds of events are
presumably detectable for the first two signatures, and models
with tens of events for the rest. The same-sign dilepton channel
has smaller backgrounds: even a handful of clean events may
constitute a signal.

These models, which are better motivated from the string theory
point of view than mSUGRA, offer far better prospects for
interesting physics at the Tevatron than even the most optimistic
unified scenario. This is in large part due to the much lighter
gluino in these models for the same value of the Higgs mass.
Considering these models in the context to supergravity was
essential to recognizing this result -- only supergravity can
truly take us from compactification to calorimeter.

%
%
%
%
%

\end{document}